\documentclass{aa}
\usepackage{epsfig}
\epsfclipon
\begin{document}

   \thesaurus{12     % Cosmology.
              (12.07.1;  % Gravitational lensing
               11.17.3;  % quasars: general
               11.17.4)} % quasars: individual:
   \title{Detection of the lensing galaxy in HE
2149$-$2745\thanks{Based on observations made at the European
Southern Observatory, La Silla. Chile}}

   \author{Sebastian Lopez 
          \and
          Olaf Wucknitz
          \and
          Lutz Wisotzki
          }

   \offprints{S. Lopez}

   \institute{
              Hamburger Sternwarte, Universit\"at Hamburg, Gojenbergsweg 112, 
              21029 Hamburg, Germany (slopez@hs.uni-hamburg.de)
             }

   \date{Received; accepted}

   \maketitle

   \begin{abstract}
We have obtained new {\it BVR}-band images of the gravitationally lensed
QSO pair HE 2149$-$2745 A,B (QSO redshift $z_e=2.033$; angular
separation $\theta_{\rm AB}=1\farcs71$), under 
1\arcsec{} seeing conditions. 
Subtraction of a scaled point-spread-function reveals an
extended object of angular size comparable to $\theta_{\rm AB}$, and $R$
magnitude of 20.4 lying between and equidistant to the QSOs. The
geometry of the system leads us to interpret this object as the
lensing agent responsible for the double QSO images. 
A fit to the galaxy surface brightness profile indicates 
an elliptical galaxy, which is further supported by the absence of
\ion{Mg}{ii} in absorption. The lensing galaxy is not detected in $B$ 
and $V$. We discuss possible lens models for this
geometry constrained by our observations and consequences for the mass
and redshift of the lens.

   \end{abstract}

\section{Introduction}
\label{sec_int}

Gravitational lensing of distant extragalactic sources (e.g.,
QSOs) by intervening large masses at low to moderate redshift (e.g.,
galaxies) has become a powerful tool of modern 
cosmology. The multiple images of lensed QSOs are
expected to be separated by a few arc seconds in the sky, implying
transverse line-of-sight (LOS) separations of up to few tens of
kpc. 
%Besides the obvious advantages multiple close LOSs have 
%for high-resolution QSO spectroscopy, 
The detection and study of the lensing agent 
provide an independent way to (i) probe the dark mass
involved in the lensing potential (e.g., Natarajan et
al.~\cite{Natarajan}); (ii) put constraints on the the Hubble
parameter $H_0$ in case of QSO variability (Refsdal~\cite{Refsdal}). 
However, the position of the lensing agent,
normally much fainter than the lensed source, is fairly difficult to
determine. 
To date, about 20 lens redshifts have been determined out of twice as
many known gravitational 
lens events.

In this letter we present new {\it BVR} 
images of the gravitationally lensed QSO pair HE 2149$-$2745 A,B
aimed at detecting the galaxy responsible for the double image. 
The QSO redshift is $z=2.033$ and the angular separation between
the QSO images is $\theta_{\rm AB}=1\farcs71$. As the QSO pair was
discovered, the 
spectroscopic and $R$-band observations by 
Wisotzki et al. (\cite{Wisotzki}) already revealed its lensing nature,
but the detection of the lensing galaxy remained uncertain.
Our new images confirm the presence of the lensing galaxy. 
The sampling is well below the small separation between the two QSO
images, so an estimate of galaxy parameters is possible. They are
discussed in 
the framework of possible lens scenarios.

\section{Observations and image analysis}
\label{sec_obs}

We observed HE 2149$-$2745 on August 8th, 1997, with SUSI on the 3.5m ESO
NTT. SUSI uses a TEK CCD with 1024$^2$ 24 $\mu$m pixels corresponding
to 0\farcs13 in a $2\farcm2\times2\farcm2$ sky field of
view. Despite a rather variable seeing several exposures could be
taken under $\sim1\arcsec$ seeing conditions. This was lucky 
given the bad
weather conditions just days before the observations. The night was
not photometric.  

We obtained science exposures in the three bands $B$ (5 exposures \`a
600 s), $V$ ( 5 \`a 400 s), and $R$ (9 \`a 300 
s). We also obtained short exposures of the photometric
standard star field PG2213-006 (Landolt~\cite{Landolt}) to check for
possible spatial variations of the
point-spread function (PSF) and to normalize our magnitudes to
the standard system. 
After bias-subtraction the images were corrected for
detector pixel-to-pixel variations using combined twilight sky
flat-fields. The effective width of a point source in
the science frames, as measured for the only well-exposed star in the
SUSI field 
(hereafter ``star 3''; cf. Fig. 1 in Wisotzki
et al.~\cite{Wisotzki}), range from 1\farcs3 to 1\farcs5 in
the $B$-band frames, from 1\farcs3 to 1\farcs4 ($V$),
and from 1\farcs0 to 1\farcs4 ($R$). Due to this quality dispersion we
decided to analyse each 
image {\em separately}. 

All {\it BVR} exposures clearly resolve the 
QSO components (hereafter A and B, respectively). 
We used the DAOPHOT II package to derive the PSF and to obtain
astrometric and photometric parameters in the frames containing the
QSOs. The PSF was 
modeled with a $\beta=2.5$ Moffat function whose parameters resulted
from the fit to star 3, allowing for empirical departures 
from the analytical form. A similar one-star fit and subsequent
subtraction of the scaled PSF in the more crowded standard-star field
demonstrated the  
spatial stability of the PSF shape, thus validating the use of just
one star in its calculation.

\section{Results}
\label{sec_res}

\subsection{$R$-band images: the lensing galaxy}
\label{sec_len}

\begin{figure}
\vspace{0cm}
\hbox{\hspace{0cm}\epsfig{figure=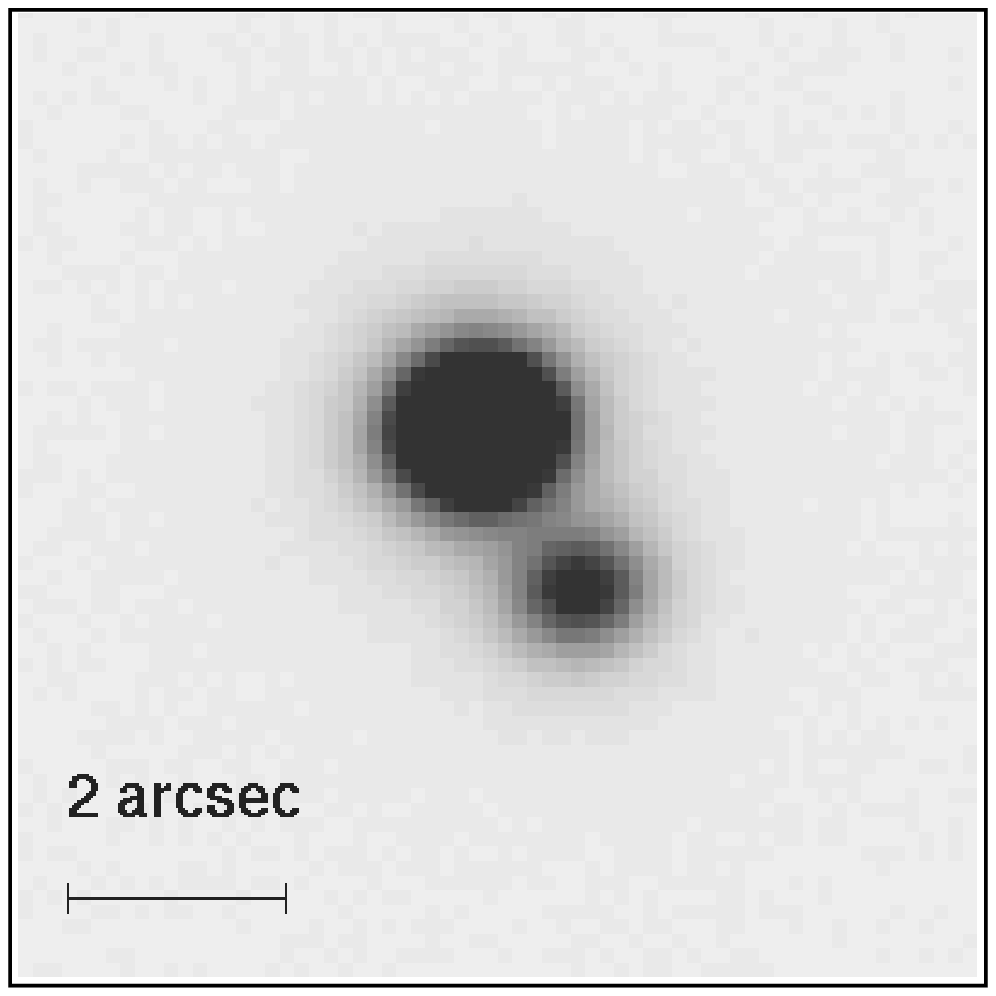,width=4.3cm}
\epsfig{figure=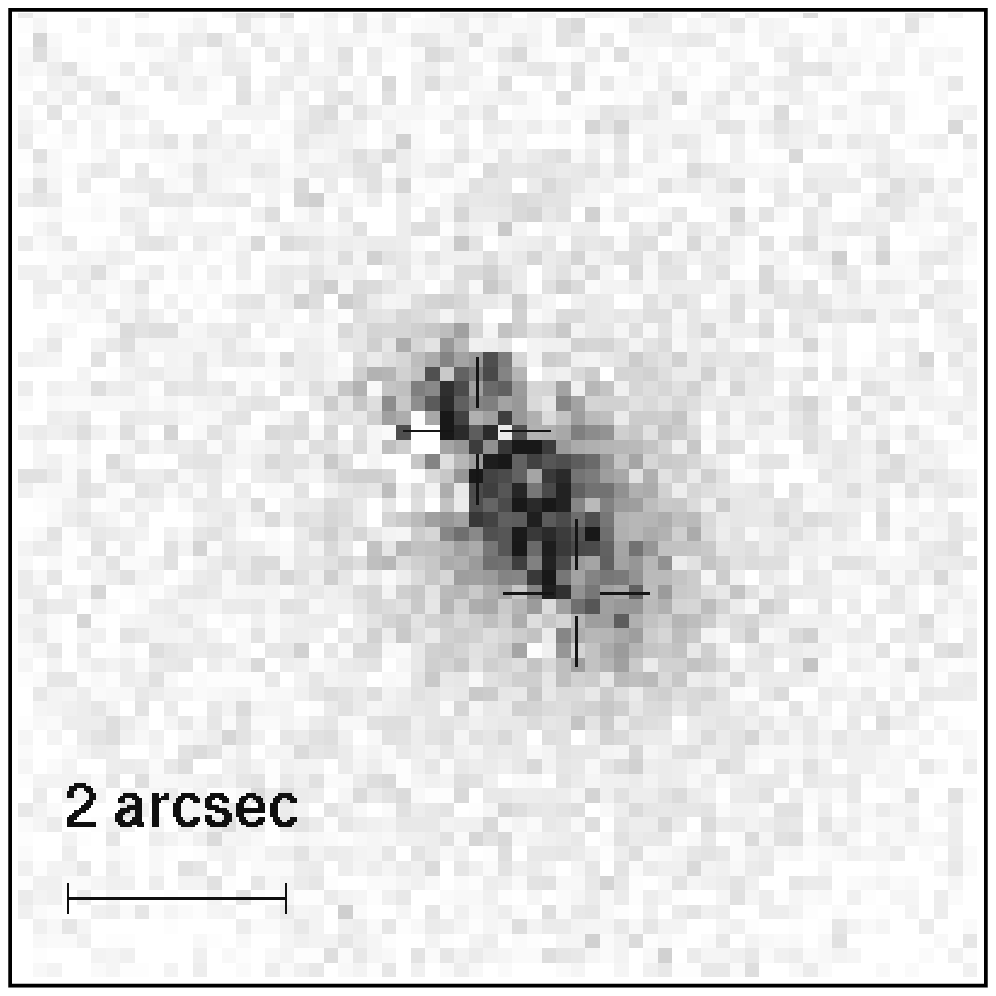,width=4.3cm}}
\vspace{0cm}
      \caption[]{{\it Left:} $R$-band 300 s exposure of HE 2149$-$2745
      A,B. North is on the top, east to the right. {\it Right:} Same
      as in left-hand panel but after scaling and subtracting the PSF
      at the positions indicated with the crosses. Two exposures with
      similar effective seeing of 1\farcs0 have been combined (600 s).
              }
         \label{fig1}
   \end{figure}

Fig.~\ref{fig1} (left) shows an $R$-band single exposure of HE 2149$-$2745 A,B
with FWHM = 1\farcs0. The subtraction of the scaled PSF at the
position of both QSO images clearly leaves a diffuse, ellipsoidal
object lying between and more or less equidistant to A and B, as is shown in
the right-hand panel. The simultaneous scaling of
the PSF overfitted the fluxes at the A and B peaks, thus leaving small
regions with relative negative residuals around both QSOs. This
is expected given the contribution from the underlying object outside
a few seeing radii; consequently, small
corrections of roughly 0.01 (A) and 0.02 (B) magnitudes 
were made to recover the lacking flux. 
The mean flux values in the central region 
deviate $15 \sigma$ from the background sky level. The stability of 
the PSF leads us to conclude that we have undoubtedly detected 
a galaxy in the LOSs to HE 2149$-$2745 A and B, most probably
the gravitational lens. 

Simple aperture photometry yields $R_{\rm G}=20.35\pm0.20$ for the galaxy.
The error bar reflects the tolerance range allowed by the fine-tuning
described in the above paragraph, and includes the zero-point 
uncertainty. For the sum of both QSO components we find 
$R_{\rm A+B}=16.30\pm0.04$. These magnitudes are based on the 
photometric results by Wisotzki et al. (\cite{Wisotzki}) for star 3
(assuming it has not varied), 
and consider a color-term correction of 0.05 which reduces them to
the Landolt (\cite{Landolt}) system\footnote{The $B$ magnitude for
star 3 given in Wisotzki et al. (\cite{Wisotzki}) was erroneous and must be
corrected to $B=16.14$}. 

The relatively deep $R$-band images reveal many red non-stellar
objects  in the SUSI field of view, but no evident
overdensity is observed close to the lensing galaxy 
to a limiting magnitude of $R \sim 23$. However, we cannot exclude the 
possibility that the galaxy belongs to a cluster since the small field
of view implies transverse distances still compatible with cluster 
sizes. In particular, two of the objects, not
much fainter than 20.4, lie within $12\arcsec$ of the galaxy position.
It remains therefore
unclear whether the galaxy belongs to a cluster or not.  

%----------------------------------------------------------- S_vib
%______________________________________________________________
%______________________________________________________________

   \begin{table}
      \caption[]{Photometry and astrometry of HE 2149$-$2745 A,B
      and the lensing galaxy.}
         \label{tab1}
      \[
%         \begin{array}{p{0.5\linewidth}rrr}
         \begin{array}{lrrrr}
            \hline
            \noalign{\smallskip}
         &\multicolumn{1}{c}{\rm A}&\multicolumn{1}{c}{\rm B}&m_{\rm A}-m_{\rm B} &\multicolumn{1}{c} {\rm Galaxy}    \\
            \noalign{\smallskip}
            \hline
            \noalign{\smallskip}
  x        & 0.00\pm0.03 &0.90\pm0.03 &&  0.50\pm0.08\\
  y        & 0.00\pm0.03 &-1.45\pm0.04&&  -0.73\pm0.07\\
            \noalign{\smallskip}
            \hline
            \noalign{\smallskip}
  B     & 17.13\pm0.03&18.70\pm0.03&1.57\pm0.04&\ga22.2^{\rm a}  \\ 
  V     & 16.82\pm0.06&18.39\pm0.06&1.57\pm0.08&\ga21.7^{\rm a}    \\ 
  R     & 16.61\pm0.02&18.20\pm0.02&1.59\pm0.03&20.35\pm0.20  \\ 
            \noalign{\smallskip}
            \hline
         \end{array}
      \]
\begin{list}{}{}
%\item[$^{\mathrm{a}}$] Mean position of first half-intensity
%isophotes.
\item[$^{\mathrm{a}}$] Approximate $3 \sigma$ magnitude limits.
\end{list}
   \end{table}
%---------------------------------------------------------------------

Table~\ref{tab1} also gives the calibrated magnitudes of both QSO
components. 
%Errors do not consider the zero-point scale
%uncertainty. 
As the $V$ magnitude of star 3 is not available we assumed
$V-R=0.46\pm0.06$ using a selected sample of stars
with similar $B-R$ colors. We note that 
the magnitude difference between A and B is consistent with a
flux ratio A to B of $f=4.3$ for all three band filters. 
%Both QSO components seem to have become brighter in $B$ by $\sim0.2$ and
%fainter in $R$ by $\sim0.1$ magnitudes,
%with respect to the first observations two years before. According to
%the measurement errors, this is indicative of intrinsic QSO
%spectral variability, but spectrophotometric observations are needed to
%confirm  this.

\begin{figure}
      \vspace{0cm}
\hspace{0cm}\epsfig{figure=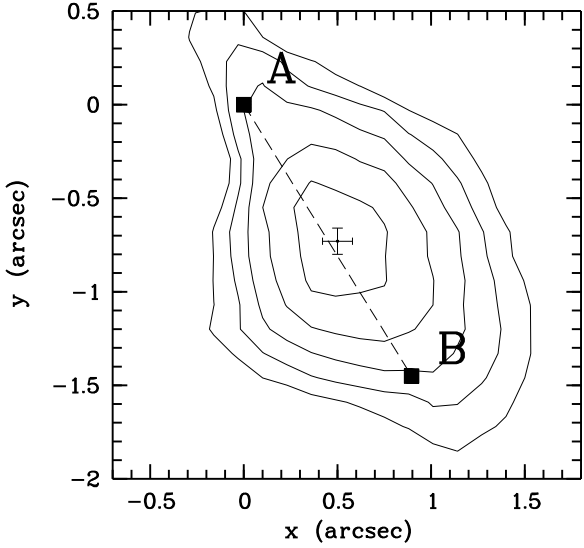,width=7.8cm}
\vspace{0cm}
      \caption[]{
       Isophote contours of the lensing galaxy in intervals of 
       $0.2$ magnitudes. The lowest contour level is at $7 \sigma$
       above the sky background.
              }
         \label{fig3}
   \end{figure}

Fig.~\ref{fig3} shows  isophotal contours of the lensing galaxy. The
isophotes are separated by $0.2$ magnitudes, and the lowest contour
level is approximately at $7 \sigma$ above the background count. Note
the distorted region near QSO A, probably due to photon-shot noise
near the A peak (see below). The positions of A and B are marked
with filled squares, since their errors are too small to appear at
this scale. The angular separation between A and B is 
$\theta_{\rm AB}=1\farcs71\pm0\farcs07$, and the position angle is
${\rm PA}_{\rm AB}=29.5\degr\pm1.0\degr$. 
The position of the galaxy was derived 
using the isophotes at brighter levels than half the peak 
intensity: $x=0\farcs50\pm0\farcs08$, and $y=-0\farcs73\pm0\farcs07$ 
relative to A. There is a small misalignment
between this position and the QSO images, but alignment is 
possible within the measurement uncertainties. Its orientation coincides
surprisingly well with the line joining A and B ($\overline{\rm AB}$),
with an ellipticity of  
$\epsilon\equiv(a-b)/(a+b)\sim0.33$. The galaxy center lies
nearly equidistant to both QSOs, with $\theta_{\rm AG}=0\farcs88$
and $\theta_{\rm BG}=0\farcs83$.

%----------------------------------------------------------- 
%______________________________________________________________

\begin{figure}
\hspace{0cm}\epsfig{figure=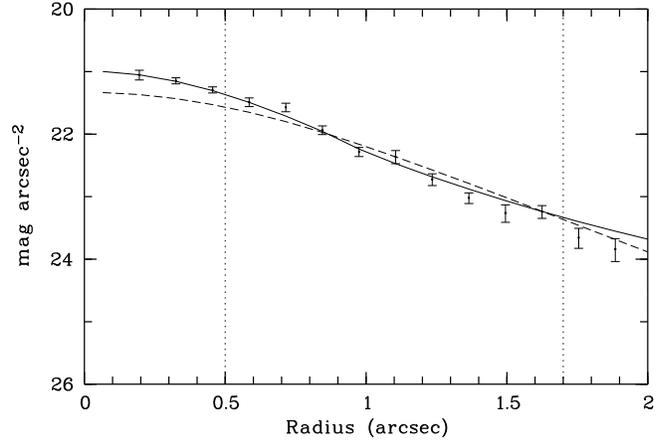,width=8.6cm}
      \caption[]{Surface brightness profile of the lensing galaxy.
      {\it Solid line:} $r^{1/4}$ law fitted
      to the data within $0\farcs5\leq r\leq1\farcs7$, and convolved with
      the PSF.{\it Dashed line:} Same for an exponential disk. 
      The fitting range is indicated by the vertical dotted lines. 
                    }
         \label{fig4}
   \end{figure}

The good spatial sampling of our data allows an estimate of the galaxy
parameters. A $180\degr$ section aligned with the south semimajor axis
was used to fit isophotal ellipses to the flux values. 
Fig.~\ref{fig4} shows the averaged surface brightness profile (error
bars). 
A de Vaucouleurs law and an exponential disk were  fitted
to the profile within $0\farcs5\leq r \leq 1\farcs7$, thus including
only values at the $> 3\sigma$ level but not significantly affected by the
seeing smoothing. The final fit models 
result from re-constructing the isophotal ellipses with the 
fitted profiles and smoothing them with the PSF. We see that the data
are better modeled by a $r^{1/4}$ law ($\chi^2_\nu=0.3$) than by an
exponential disk ($\chi^2_\nu=2.2$), 
especially at the core.  
Furthermore, images taken under better seeing conditions
should accentuate the concave shape of the observed profile. 
We take this result, though cautiously, as evidence for an elliptical galaxy.
%
%
%The data points for radii larger than
%$\sim1\farcs7$ indicate a surface brightness somewhat lower than 
%expected, but the uncertainties become larger here 
%and the measured points may not represent the true galaxy
%profile. At smaller radii, we expect the seeing smoothing to be
%significant up to length scales of $\sim 0\farcs5$, which is also
%observed. 
%
%In any case, the present data allows to exclude a disk
%model for the galaxy as the length scales we are dealing with are
%comparable to 
%the seeing; consequently, 
%

\subsection{$B$ and $V$ band images}
\label{sec_ana}

The galaxy is not detected in the PSF-subtracted $B$ and $V$ band
images. However, the region
surrounding QSO image A is slightly overfitted at both sides of
$\overline{\rm AB}$. Such residuals are not 
observed for B. This ``symmetric''
overfit, though not significant ($\la 2\sigma$ in $B$ and only
marginal in $V$), suggests A might be composed of two or
more fainter point sources lying on $\overline{\rm AB}$. 
We investigated this possibility (cf., Bade et al.~\cite{Bade}; Burud
et al.~\cite{Burud}) in the $B$-band, as one expects here 
less contamination by a hypothetical foreground galaxy (in spite of
the fact that the PSF is broader in the blue). 
However, attempts to re-fit A with two sources of nearly half its 
intensity failed at recovering the background level. 
It is difficult to establish on the basis of the present data
what causes the low-quality fits in the $B$ images. From an
observational point of view, splitting of QSO image A in very close
sub-components  cannot be ruled out, but an underlying object could 
also contribute to slightly distort the QSO images. 
An explanation of this must await better quality data.

To put upper limits on the galaxy $B$ and $V$ brightness a
variance frame was created which considers photon 
statistics (dominated by the QSO fluxes), readout noise, and the
uncertainties introduced by background subtraction. 
Integration of the variances in a 3\farcs5 radius aperture
yields  detection limits for $B$ 
and $V$. In addition, we consider the fraction of possible galaxy
light hidden in the QSO seeing disks and hence subtracted with the
PSF. Analysis of $R$ images with similar seeing as in $B$ and $V$
shows that this fraction can be as large as 0.7 ($B$) and 0.5
($V$). The estimated ($3\sigma$) detection limits are listed in 
Table~\ref{tab1}.

%----------------------------------------------------------- 
%
%______________________________________________________________

\section{Discussion}
\label{sec_dis}

\subsection{Redshift of the galaxy}

The redshift of the lensing galaxy can in principle be constrained by
its color  
if we know its morphology. Besides the surface-profile fits, further
evidence that we are observing an elliptical galaxy in front of HE
2149$-$2745 comes from the optical spectra of QSO A: no \ion{Mg}{ii}
absorption system is detected down to an (observed) equivalent width
limit of $\sim 0.3$ \AA{} ($3 \sigma$). Given the small impact parameter 
($\sim 7$  $h_{50}^{-1}$kpc at $z=1$), this result almost excludes a
disk-like galaxy in the foreground at $z>0.35$. 

The $3 \sigma$ lower limit derived for the galaxy $V$ magnitude
(Table~\ref{tab1}) puts a lower limit of $V-R\ga1.4$. This 
is the $V-R$ color an E galaxy would have at a redshift of $\sim0.3$ or
larger (using spectral energy distributions observed at $z=0$ from 
Coleman, Wu \& Weedman~\cite{Coleman}). 
A bound consistent with this redshift is obtained from the $B-R$ color
(Bressan, Chiosi \& Fagotto~\cite{Bressan}). The implied absolute
luminosity for $z_{\rm G}=0.3$ is $M_R=-21.5$ (considering 
$K$-corrections, $H_0=50$ km~s$^{-1}$~Mpc$^{-1}$, and 
$q_0=0.5$), i.e., very close to $M^{\ast}$. An upper limit for 
$z_{\rm G}$ is difficult to establish, but for the observed $R_{\rm
G}$ the expected 
luminosity beyond $z\sim0.5$ becomes too large to be real. We thus
arrive to $0.3\la z_{\rm G} \la 0.5$.

\subsection{Lens models}

Because of the symmetry of the
mass distribution expected for any regular galaxy, the deflection
angle and the amplification should be very similar for the two
images of HE 2149$-$2745. This is also true if we include an external
shear. To explain the flux ratio of 4.3, the dependence of the
amplification on the positions has to be very strong. This can be
achieved if the images are located near a critical curve, implying high
amplifications.

Given the required sensitivity of the models for small changes of the
positions and the small number of constraints, a  maximum likelihood
model fitting is not appropriate for this system; instead, we use an
analytical approach to find the possible    
model parameters considering the measurement uncertainties.

We use a singular isothermal elliptical mass distribution (SIEMD) as
given by Kassiola \& Kovner (\cite{Kassiola}).
As can be seen from Fig.~\ref{fig3},
the images are almost exactly 
located on the major axis of the galaxy. To simplify the calculations,
we use the line $\overline\mathrm{AB}$ as the major axis and project
the center of the 
galaxy onto this line.
We further include an external shear $\gamma$, whose
source has to be located on the major or minor axis to
be in agreement with the observed image positions. 
As observational parameters, we use the ratio of distances of the
images from the center of the galaxy $f_x =\theta_{\rm BG}/\theta_{\rm
AG}$ (nearly unity) and the amplification ratio 
$f=4.3$. In addition, we force the two images to have different
parity, which is a necessary condition to exclude the existence of
more than two images. Even non-singular models (PIEMD) rule out the
possible splitting of A in the radial direction.

On the main axis, the lens equation and the amplification for the
SIEMD model with external shear read
\begin{eqnarray}
x_\mathrm{s} &=& (1-\gamma)\,x-\alpha_0\frac{1-\epsilon^2}{2\sqrt\epsilon}
\arctan\frac{2\sqrt\epsilon}{1-\epsilon}\,\mathrm{sign}\,x \nonumber\quad ,\\
M_x^{-1} &=&
(1-\gamma)\left(1+\gamma-\frac{\alpha_0\,(1+\epsilon)}{|x|}\right)
\nonumber  \quad .
\end{eqnarray}

\begin{figure}
\vspace{0cm}
\hspace{0cm}\epsfig{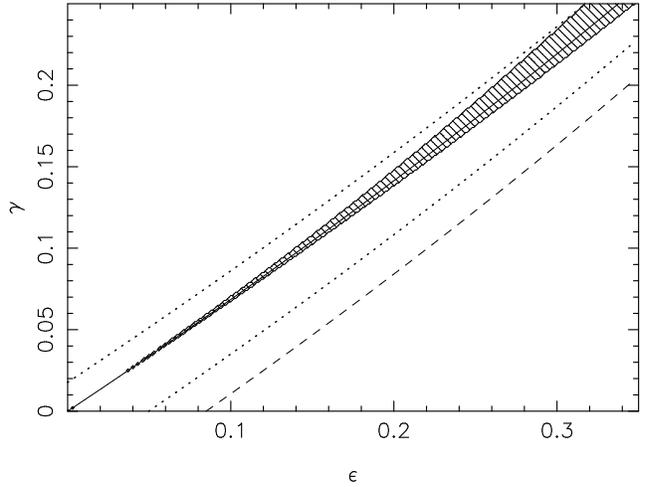}
\vspace{0cm}
      \caption[]{Model parameters $\epsilon$ and $\gamma$ consistent
        with the observations. Models with more than two images are
        shown hatched. The dotted lines are $1\sigma$ bounds of the
        measurements, the dashed line is a lower limit for $\gamma$
        independent of $f_x$. The solid line shows a symmetrical
        configuration with $f_x=1$.
              }
         \label{epsgam}
   \end{figure}

Since a degeneracy in the models prevents the independent determination of
$\epsilon$ and $\gamma$, we use the two above equations to define the
parameter $E$, 
\begin{displaymath}
E \equiv \frac{1+\gamma}{1-\gamma} \frac{1-\epsilon}{2\sqrt\epsilon}
\arctan\frac{2\sqrt\epsilon}{1-\epsilon} 
= \frac{(1+f_x)(1+ff_x)}{2f_x(1+f)} \quad .
\end{displaymath}
With our data, we get $E=0.99 \pm 0.05$ 
($1\sigma$). The uncertainty in $E$ is dominated by 
the errors in the galaxy position. If the
mass distribution has the same ellipticity as the light
($\epsilon=0.33$), an external shear of $\gamma_\mathrm{min}=0.21$ is
needed to keep $E$ inside the $1\sigma$ bounds. Because of the absence
of very close galaxies or a rich cluster in the field, we do not expect
such a large shear. For an ellipticity of $\epsilon=0.2$, the minimal shear
decreases to $\gamma_\mathrm{min}=0.11$. Fig.~\ref{epsgam} shows the
possible parameters consistent with the measured positions and the
flux ratio.
Even for different parity of the images some very
symmetrical models lead to more than two images (up to {\em eight} 
images are possible for SIEMD+shear models).

Only rough estimates for the absolute amplifications can be determined
from the observations. For a best-fit 
model, we get $M_\mathrm{A}\approx 70$. Considering the errors,
a lower limit for $|M_\mathrm{A}|$ of $27$ (68\% confidence) can be
obtained.

%------------------------------------------------------------------------
%---------------------------------------------------------------------------

To estimate the mass and velocity dispersion of the galaxy, we use a
spherical model ($\epsilon=0$). For lens redshifts of
$z_{\rm d}=0.3~(0.5)$, the mass inside the Einstein radius is
$M=1.5~(2.4)\times10^{11}\,h_{50}^{-1}M_\odot$, and the
velocity dispersion
$\sigma_v=205~(230) \mathrm{\,km\,s^{-1}}$ ($\Omega=1$, 
$\lambda=0$). The implied mass-to-light ratio is $M/L_R=5~(2)$ in
solar units. 

The expected order of magnitude for the time delay is
about weeks. A better estimate must await more stringent
constraints on the galaxy position. Given the geometry of the system,
off-center spectroscopy of the galaxy should be possible from the
ground under excellent seeing conditions, or with STIS onboard the
{\it HST}.

\begin{acknowledgements}
We thank K. Jahnke for computing the surface brightness profile.
\end{acknowledgements}

\end{document}